# Ubiquitous Interplay between Charge Ordering and High-Temperature Superconductivity in Cuprates

Eduardo H. da Silva Neto[1*†], Pegor Aynajian[1*‡], Alex Frano[2,3], Riccardo Comin[4], Enrico Schierle[3], Eugen Weschke[3], András Gyenis[1], Jinsheng Wen[5], John Schneeloch[5], Zhijun Xu[5], Shimpei Ono[6], Genda Gu[5], Mathieu Le Tacon[2], and Ali Yazdani[1§]

[1]Joseph Henry Laboratories and Department of Physics, Princeton University, Princeton, New Jersey 08544 USA
[2]Max-Planck-Institut für Festkörperforschung, Heisenbergstrasse 1, D-70569 Stuttgart, Germany
[3]Helmholtz-Zentrum Berlin für Materialien und Energie, Albert-Einstein-Strasse 15, D-12489 Berlin, Germany
[4]Department of Physics & Astronomy, University of British Columbia, Vancouver, British Columbia V6T 1Z1, Canada
[5]Condensed Matter Physics and Materials Science, Brookhaven National Laboratory, Upton, NY 11973 USA
[6]Central Research Institute of Electric Power Industry, Komae, Tokyo, Japan

[*]These authors contributed equally to this work.
[†]Present address: Quantum Matter Institute, University of British Columbia, Vancouver, British Columbia V6T 1Z4, Canada
[‡]Present address: Department of Physics, Applied Physics and Astronomy, Binghamton University, Binghamton, NY 13902, United States
[§]To whom correspondence should be addressed. Email: yazdani@princeton.edu

**Besides superconductivity, copper-oxide high temperature superconductors are susceptible to other types of ordering. We use scanning tunneling microscopy and resonant elastic x-ray scattering measurements to establish the formation of charge ordering in the high-temperature superconductor $Bi_2Sr_2CaCu_2O_{8+x}$. Depending on the hole concentration, the charge ordering in this system occurs with the same period as those found in Y-based or La-based cuprates, and displays the analogous competition with superconductivity. These results indicate the similarity of charge organization competing with superconductivity across different families of cuprates. We observe this charge ordering to leave a distinct electron-hole asymmetric signature (and a broad resonance centered at +20 meV) in spectroscopic measurements, thereby indicating that it is likely related to the organization of holes in a doped Mott insulator.**

Understanding the mechanism of superconductivity and its interplay with other possible spin or charge organizations in high transition temperature ($T_c$) cuprate superconductors remains one of the most significant challenges in condensed matter physics. The observation in La-based cuprates of a suppression of $T_c$ near x = 1/8 doping (hole/Cu) coinciding with the organization of charge into stripe-like patterns with four lattice constants (4a) periodicity, provided early evidence that charge ordering (CO) competes with superconductivity (*1, 2*). Recently, x-ray scattering experiments on highly ordered Y-based cuprates have discovered that in the same range of hole concentration (near x = 1/8) there is a competing charge organization with an incommensurate ordering vector (≈ 3.3a periodicity) (*3, 4*). These experiments raise the question of whether there is a universal charge ordering mechanism common to all underdoped cuprates [though crystalline structures in each compound may modify details of the resulting ordering patterns (*5, 6*)]. If there is a connection between the observed COs in different compounds, what is the mechanism by which this phenomenon competes or is intertwined with high-$T_c$ superconductivity (*7*)? Do strong correlations, such as those present as a result of proximity to the Mott insulator, play a role in the CO state and its interplay with superconductivity?

To address these questions, we have carried out scanning tunneling microscopy (STM) and spectroscopy, as well as resonant elastic x-ray scattering (REXS) measurements, on underdoped $Bi_2Sr_2CaCu_2O_{8+x}$ (Bi-2212) samples. The Bi-2212 system is the only material system among the cuprates for which there is detailed spectroscopic information on the Fermi surface and the occupied electronic states as a function of temperature and doping, obtained from angle-resolved photoemission spectroscopy (ARPES) performed over the last two decades (*8-*

*10*). Yet, ARPES measurements have not shown any evidence of band folding associated with CO along the Cu-O bond direction in this system (*9, 11, 12*). STM studies have long reported evidence of spatial modulation of electronic states in Bi-based cuprates (*13-19*), but evidence for CO in these experiments is often obfuscated by the disorder-induced energy-dispersive quasiparticle interference (QPI) effect (*19-21*). The analysis of STM conductance maps over a range of energies has been used to provide evidence for fluctuating charge organization in Bi-2212 below T*, with strongest enhancement near 1/8 doping – the same doping range where charge organizations have been seen in Y- and La-based cuprates (*19*). However, it remains a challenge to clearly separate signatures of CO from those of QPI in STM experiments, and more importantly, to corroborate the surface-sensitive measurements with bulk sensitive experiments. Here we provide high-resolution energy-revolved STM spectroscopy experiments that clearly distinguish between QPI and CO features as a function of doping and temperature; this CO signature is also detected in our temperature-dependent REXS measurements.

Figure 1A shows the geometry of the combined experimental approach for examining CO in underdoped Bi-2212 samples, where we contrast the discrete Fourier transform of STM conductance maps with bulk scattering results from REXS. A typical example of an STM conductance map is shown in Fig. 1B for a UD45 (underdoped Bi-2212, $T_c$ = 45 K) sample at 30 K. The discrete Fourier transform of the conductance map (Fig. 1C-D) shows a strong peak of wavevector $q$ = ($\pm\delta$, 0)$2\pi/a$ and (0, $\pm\delta$)$2\pi/b$, and $\delta \approx 0.3$, along the Cu-O bond directions, corresponding to the real space modulations in Fig. 1B. Figure 1E-F shows the results from REXS measurements (incident photon energy of 931.5 eV, resonant with the Cu $L_3$-edge) on the same sample, as a function of the component of momentum transfer along the a direction (see Fig.

1A and (*22*)). Notably, the low temperature REXS measurement (Fig. 1E) shows a peak at the same incommensurate wavevector ($\delta \approx 0.3$) as those in the discrete Fourier transform of the STM conductance maps (Fig. 1C-D), thereby establishing, that the STM modulations on the surface of Bi-2212 sample are in fact due to a CO that can also be detected in the bulk (see Ref. (*23*) for similar finding on Bi-2201).

To understand why such a charge ordering phenomenon has remained undetected in ARPES studies and to reveal a key spectroscopic characteristic of this ordering, we examine the energy dependence of STM conductance maps. Figure 2 shows the energy evolution of the features in the discrete Fourier transform of the conductance maps, along the Cu-O bond direction, measured at 30 K for a range of doping, from optimally doped (OP91) to strongly underdoped (UD15) samples. In the optimally doped sample (Fig. 2A) at temperatures well below $T_c$, the energy-wavevector structure along the Cu-O bond direction shows no sign of CO. Instead, it shows an energy-dispersing particle-hole symmetric wavevector originating from disorder-induced scattering interference of superconducting Boguliobov-de Gennes quasiparticles (BdG-QPI) (*13, 18, 19, 24*). Reducing the doping towards the underdoped regime (Fig. 2A-G), we find that the BdG-QPI features are systematically weakened, whereas a separate non-dispersive modulation with a relatively sharp wavevector appears and strengthens peaking near UD35 (Fig. 2F). This wavevector (see for example $\delta = 0.3$ in Fig. 2E-G) corresponds to the CO wavevector we find in the REXS measurements (see Fig. 1E and fig. S7). The widths in momentum of this CO peak (full width at half maximum, $2\Gamma \approx 0.04$ rlu), which agree well with our REXS measurements ($2\Gamma \approx 0.05$ rlu, or equivalently 20a in real space), are larger than those recently observed in the Y-based cuprates (*3*) and indicate a rather short-range order. Our energy resolved STM conductance maps, further show that these CO modulations only appear over a range of energies (0 to 50 meV) above the chemical potential, a behavior not expected in a conventional CDW order. Despite being broad in energy, we find the intensity of the CO feature to be centered at approximately +20 meV above the chemical potential at all doping levels (Fig. 2H). The particle-hole asymmetry of the CO feature in the STM data clearly distinguishes it from the BdG-QPI signals and more importantly explains the absence of such CO features in the ARPES studies.

The CO wavevectors extracted from our STM measurements show distinct similarities to other families of the cuprates. Figure 2H shows that for a hole concentration of $x < 0.1$, the incommensurate CO wavevector $\delta \approx 0.3$ matches the CDW observed in the Y-based cuprates, whereas for $x > 0.1$, the nearly commensurate wavevector $\delta \approx 0.25$ is very similar to that found in the stripe phase of La-based cuprates. Though an incommensurate $\delta$ and its decrease with doping are expected from a Fermi surface nesting mechanism, the narrow doping range over which the jump in $\delta$ occurs in Bi-2212 may be an indication of possible competition between two different stable forms of CO in the cuprates. More broadly though, the fact that CO in Bi-2212, in a range of doping without any structural distortion, can be either similar to Y-based or La-based cuprates demonstrates our key point that CO is a ubiquitous phenomenon to underdoped cuprates.

Further evidence that connects measurements on Bi-2212 to other cuprates and probes the interplay between CO and superconductivity comes from examining the temperature dependence of the REXS and STM data. Figure 1F shows that the CO signature in our REXS measurements is first detected at relatively high temperatures. By lowering of temperature, the intensity of the CO increases only down to $T_c$, below which it gradually weakens suggesting a competition with superconductivity. More detailed signatures of the interplay between CO and superconductivity are observed in our STM measurements as a function of temperature for a UD75 sample (where temperatures much lower than $T_c$ can be accessed in our experiments). The CO signature above the chemical potential (near $\delta \approx 0.25$ for this sample) is very strong at temperatures just above $T_c$ (Fig. 3A), while being nearly absent when the sample is cooled to roughly $0.15*T_c$ (Fig. 3D). Instead, at low temperatures, as Fig. 3D shows, the STM data for this weakly underdoped sample recover the particle-hole-symmetric dispersing features that are due to BdG-QPI, consistent with a d-wave superconducting gap (*25*). This trend is analogous to the doping dependent STM data of Fig. 2, where the CO (measured at a constant temperature, T = 30 K) peaks at the UD35 sample (where T ≈ $T_c$). Together, these findings not only clearly show superconductivity wins over CO at low temperatures, but also provide key information about this competition in momentum space.

Previous analyses of impurity-induced QPI data at low temperatures, such as the data displayed in Fig. 3D, associated the observed dispersing wavevectors along the Cu-O bond direction with the scattering of BdG quasiparticles between points on the Fermi surface. Figure 3E shows a schematic Fermi surface for a Bi-2212 UD75 sample together with the different QPI wavevectors connecting different segments of the Fermi surface, starting from near the antinodes at large energies (small $\delta$) to points near the nodes at low energies (large $\delta$)(*13, 18, 24, 25*). The Fermi surface of this sample, like other underdoped cuprates, is gapped at the antinodal region due to the formation of the pseudogap at high temperatures T* [which is well above $T_c$ for this sample]. Contrasting STM data above and well below $T_c$ shows, first, that the same quasiparticles involved in the CO at high temperature are associated with superconducting pairing at low temperatures and, second, that those quasiparticles occupy regions near the ends of the Fermi arcs (Fig. 3E) (*22, 23*). The fact that such a competition occurs away from the antinodes suggests that charge organization is secondary to the formation of the pseudogap phase (*19*).

The unusual appearance of CO only when tunneling electrons into the sample from the STM tip (i.e., only for energies above the chemical potential), and its absence in ARPES studies, suggests that the nature of this organization is related to the strong electronic correlations, which have long been predicted to give rise to asymmetric tunneling in the doped Mott insulators (*26, 27*). In a simplified picture, if holes doped in a Mott insulator organize into patterns, tunneling electrons into the sample at low energies would be much easier into the hole sites as opposed to electron sites because double occupancy is energetically unfavorable. In contrast, tunneling electrons out of the sample can easily occur at any site because either electrons are already present or can hop there from a nearby site on the lattice. Whether such constraints on the tunneling processes in a Mott system can explain the broad resonance (+20 meV) we observe above the Fermi energy (which is remarkably independent of doping) requires further investigations. Overall while much remains to be understood about the underlying mechanism of charge ordering or its unusual characteristics reported here, its universality across cuprate families establishes it as a critical component for understanding the electronic properties and superconductivity of high-$T_c$ cuprates.

**ACKNOWLEDGEMENTS**

The work at Princeton was primarily supported by a grant from DOE-BES. The instrumentation and infrastructure at the Princeton Nanoscale Microscopy Laboratory used for this work were also supported by grants from NSF-DMR1104612, the NSF-MRSEC program through Princeton Center for Complex Materials (DMR-0819860), the Linda and Eric Schmidt Transformative Fund, and the W. M. Keck Foundation. Work at BNL was supported by DOE under Contract No. DE-AC02-98CH10886. The Max Planck – UBC Centre for Quantum Materials and CIFAR Quantum Materials, also supported this work. We thank P.W. Anderson, E. Abrahams, S. Kivelson, S. Misra, and N. P. Ong for fruitful discussions. We also acknowledge A. Damascelli and B. Keimer for discussions and for sharing the results of their x-ray studies on Bi-2201 prior to publication. AY acknowledges the hospitality of the Aspen Center for Physics, supported under NSF Grant No. PHYS-1066293.

**FIGURE CAPTIONS**

**Fig. 1. Charge ordering in Bi-2212 (A)** Schematic of the combined experimental STM-REXS approach. REXS experiments were performed with vertically polarized photons (σ) in a horizontal scattering geometry (*22*) and yield momentum space information corresponding to the real space modulations seen by STM (front panel), or, more directly, to the discrete Fourier transform of the STM real space data (back panel). We use the tetragonal crystal structure where *a=b*≈3.8 Å represent the nearest neighbor Cu-Cu distance. **(B)** STM conductance map (15 meV, normalized to its spatial average) on a UD45 sample measured at 30 K Inset shows the direction of the Cu-O-Cu bond direction relative to the measurement. White bar represents 140 Å. **(C)** Discrete Fourier transform of the conductance map in (**B**), where the corners of the image represent the atomic Bragg peaks at (±2π/a, 0) and (0, ±2π/b). The discrete Fourier transform is mirror-symmetrized and normalized to its average value. The color scale represents the power spectral density (PSD) normalized to the δ = 0.3 peak (blue arrow), which corresponds to the real space modulations seen in (**B**). The central peak (δ < 0.2) corresponds to long wavelength inhomogeneity of the conductance map. **(D)** Line-cut of the energy-integrated (0 to 50 meV) discrete Fourier transform along the Cu-O bond direction (red arrow in (**C**)) showing a clear peak at δ = 0.30. Dashed line represents fit to a Lorentzian plus a background (*22*). **(E)** REXS δ-scans (*22*) for positive δ at low (blue) and high (red) temperatures on the same sample as in (B), showing a clear enhancement near δ = 0.3 at low temperatures. Inset: background subtracted 10 K data, where the dashed line represents a Lorentzian fit to the data (*22*), in the same units as (**E**). (F) The REXS intensity extracted from the background-subtracted peak maxima as a function of temperature (*22*). The vertical dashed line corresponds to $T_c$ = 45 K.

**Fig. 2. Doping dependence of the STM data. (A-G)** Energy-momentum structure of the modulations seen in STM along the Cu-O bond direction, extracted from line cuts along the ($2\pi/a$, 0) direction of the discrete Fourier transforms measured at 30 K for different doping levels (*22*). **(H)** Energy and $\delta$ location of the CO feature as a function of doping, extracted from energy integrated $\delta$ cuts (*22*). The error bars in $\delta$ represent the half width at half maximum of the extracted peaks. The error bars in energy represent the standard error obtained from a least-squares fit to a Lorentzian. The green dashed line represents the energy location averaged from all dopings.

**Fig. 3. Temperature dependence of the STM data. (A-D)** Energy-momentum structure of the modulations seen in STM along the Cu-O bond direction, extracted from line cuts along the ($2\pi/a$, 0) direction of the discrete Fourier transforms for a UD75 sample measured at selected temperatures. The data show the opposite temperature dependence between the particle-hole symmetric BdG-QPI and the particle-hole asymmetric CO. **(E)** Schematic layout of the Fermi surface in Bi-2212 UD75 sample. The green segment represent the Fermi arc as determined by ARPES above $T_c$ (*22, 28*). The vertical lines (also reproduced horizontally in **A-D**) correspond to QPI wavevectors connecting the Fermi surface (consistent with ARPES, see (*22, 29*)) at different regions.

...

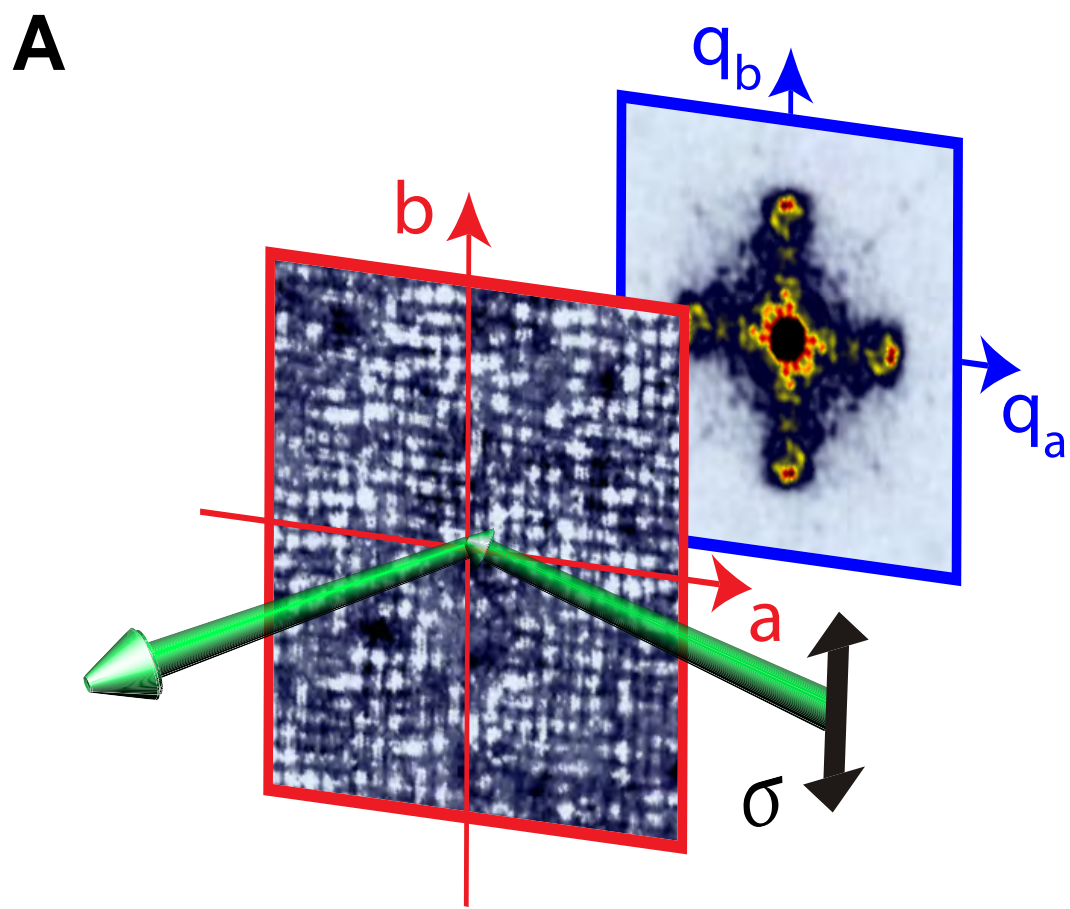
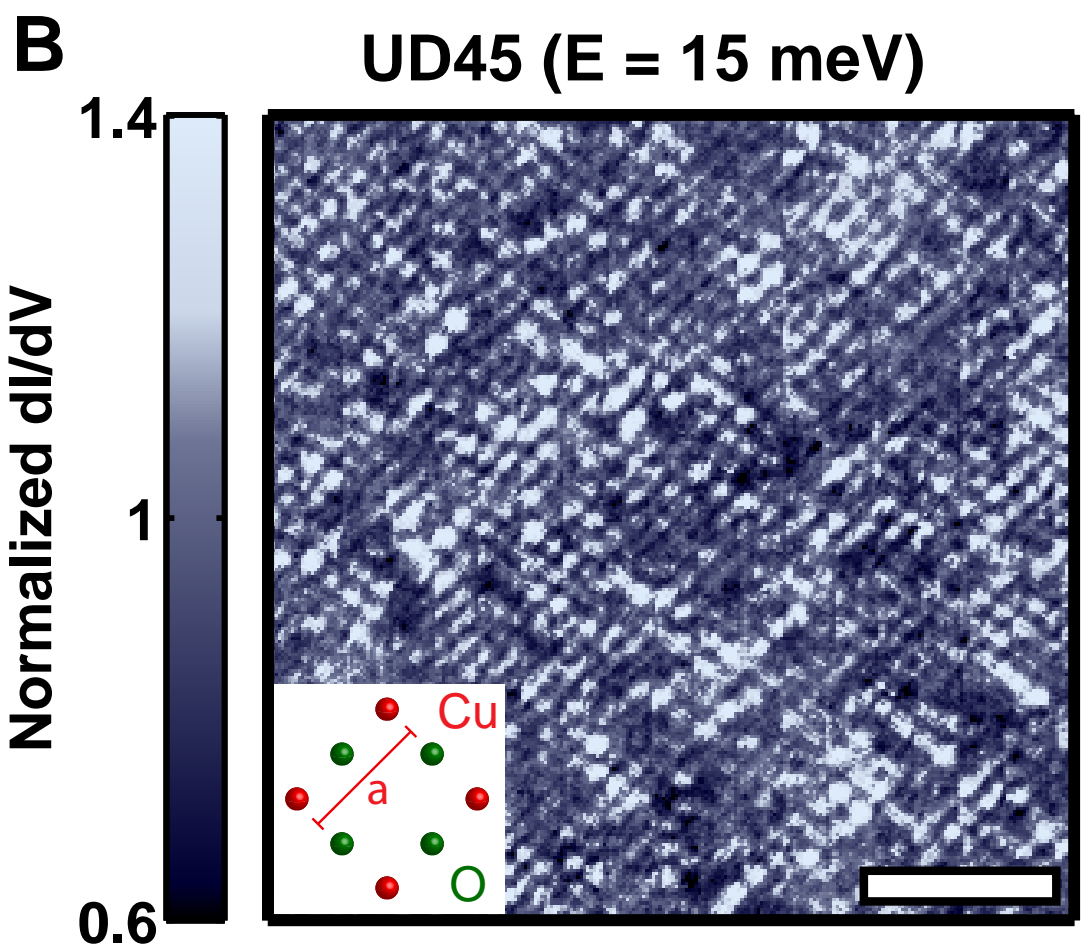
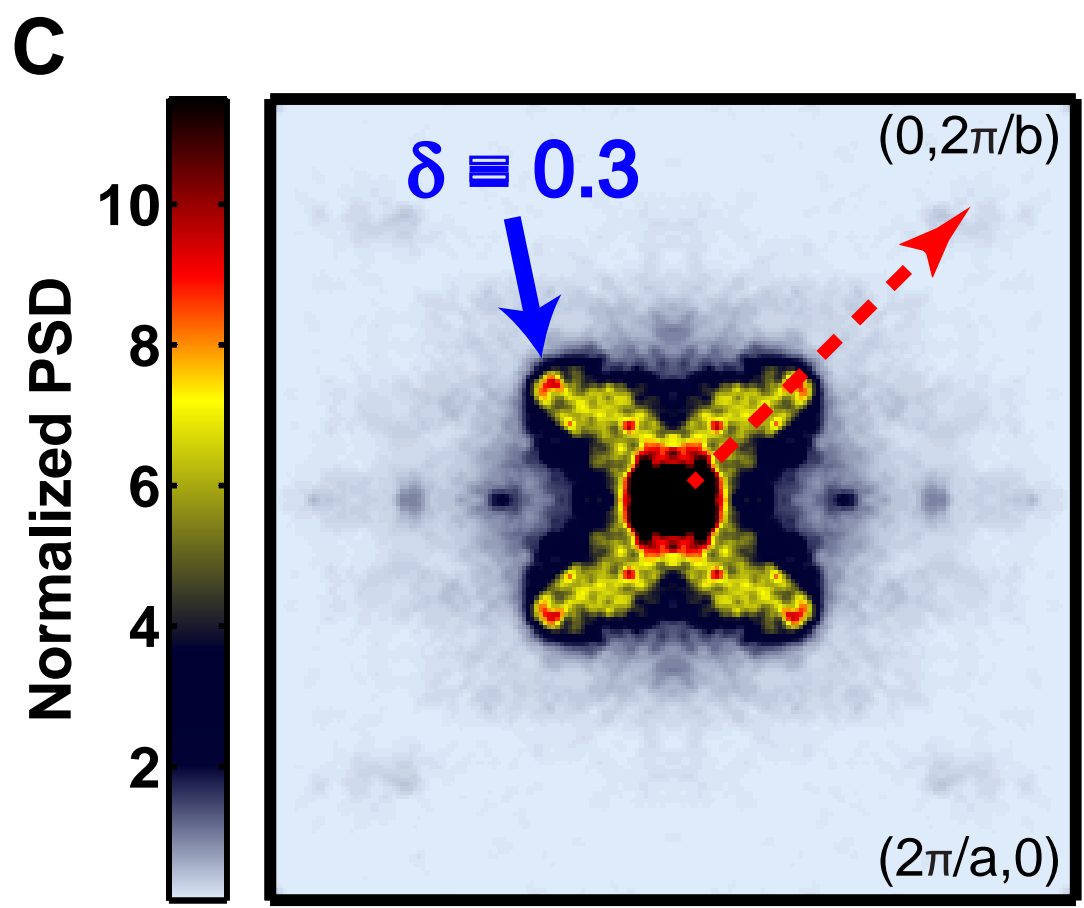
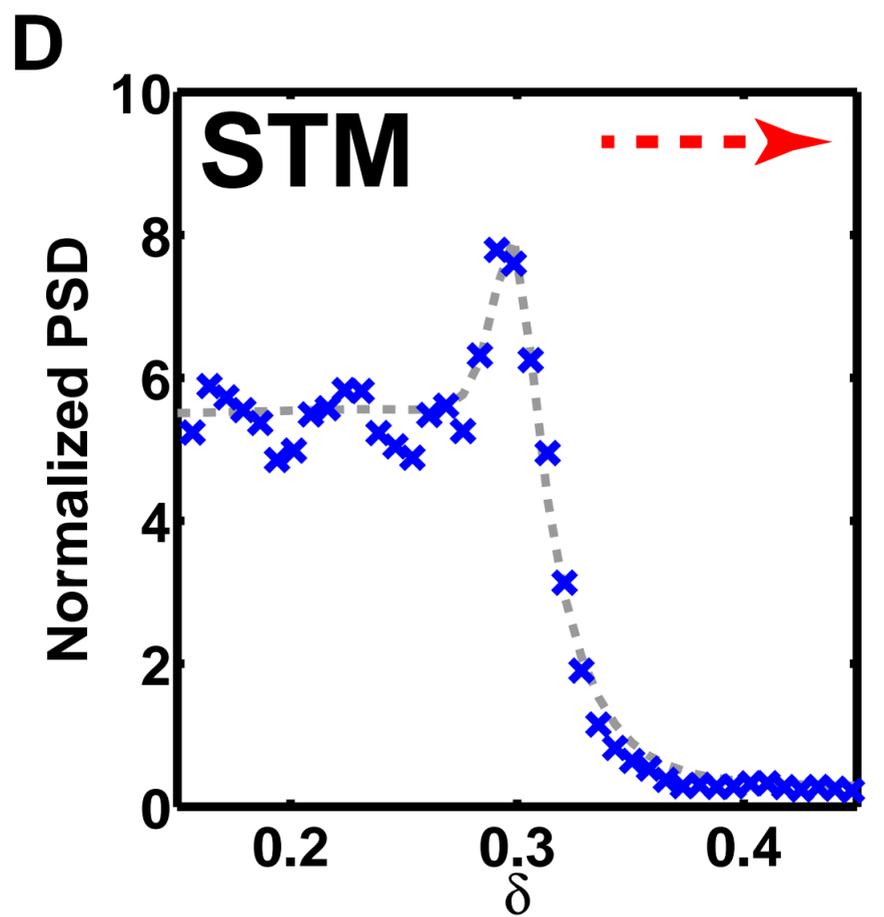
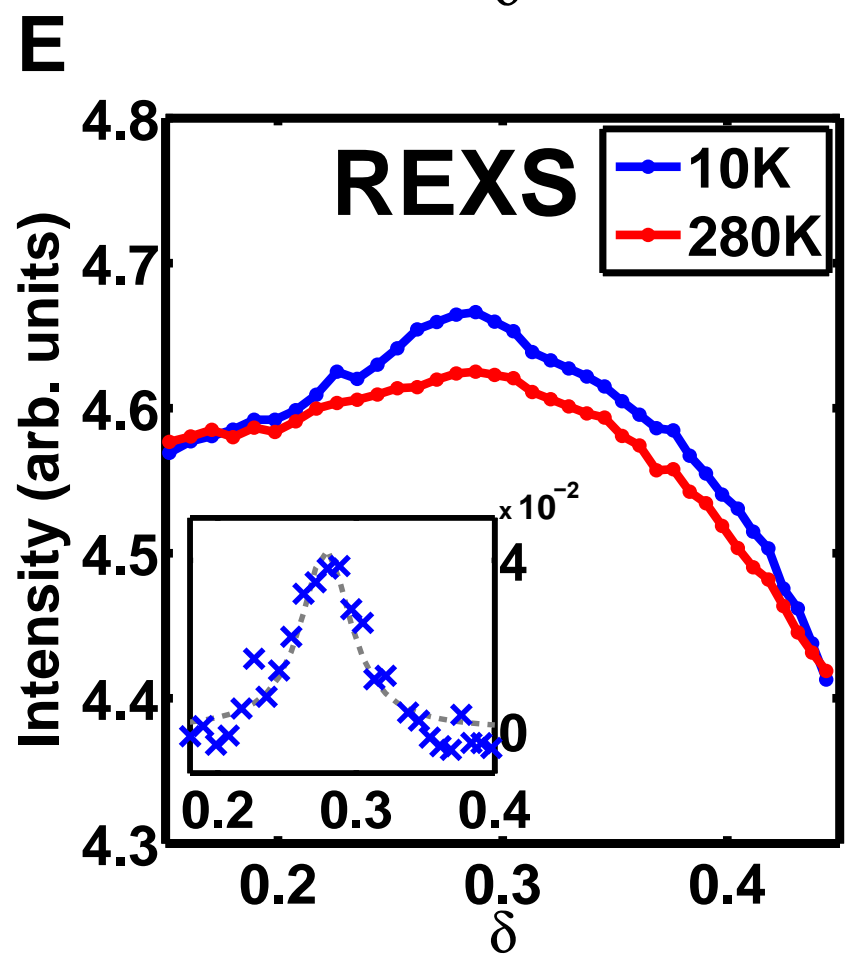
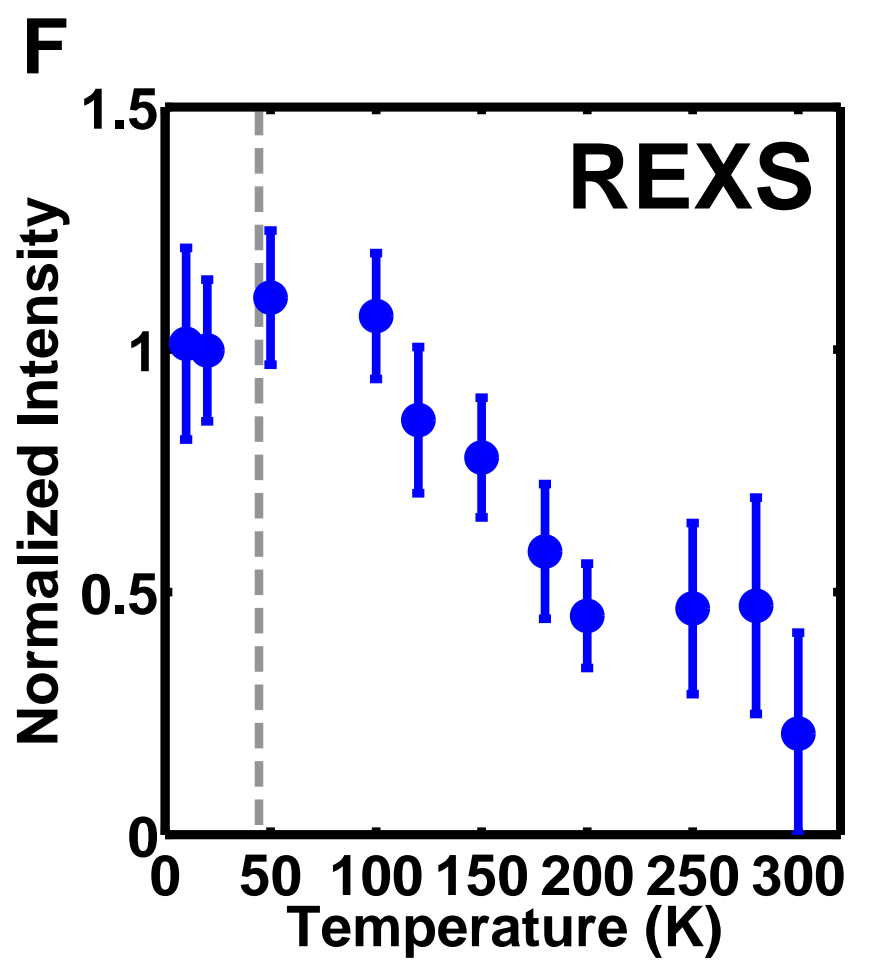

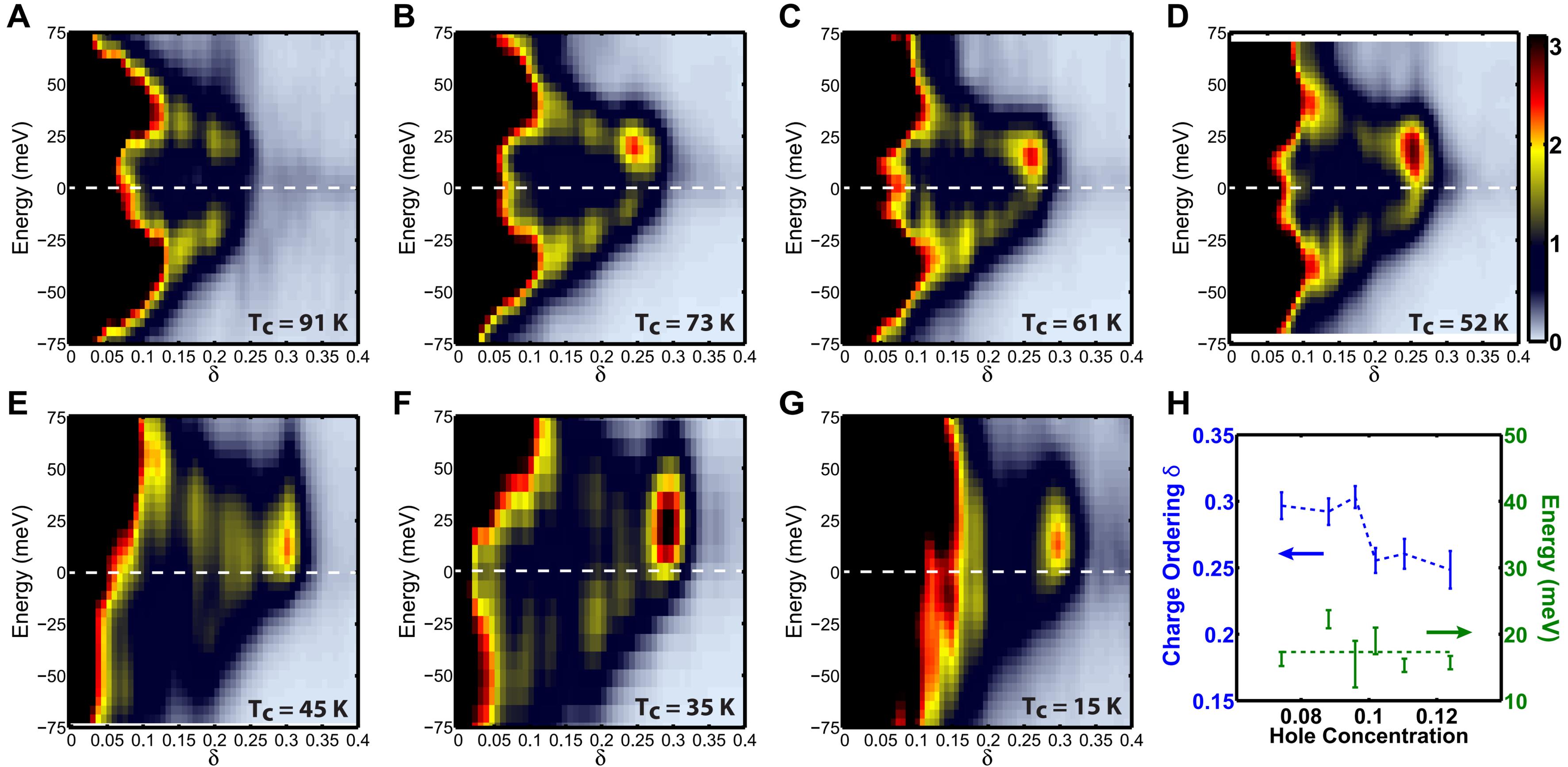

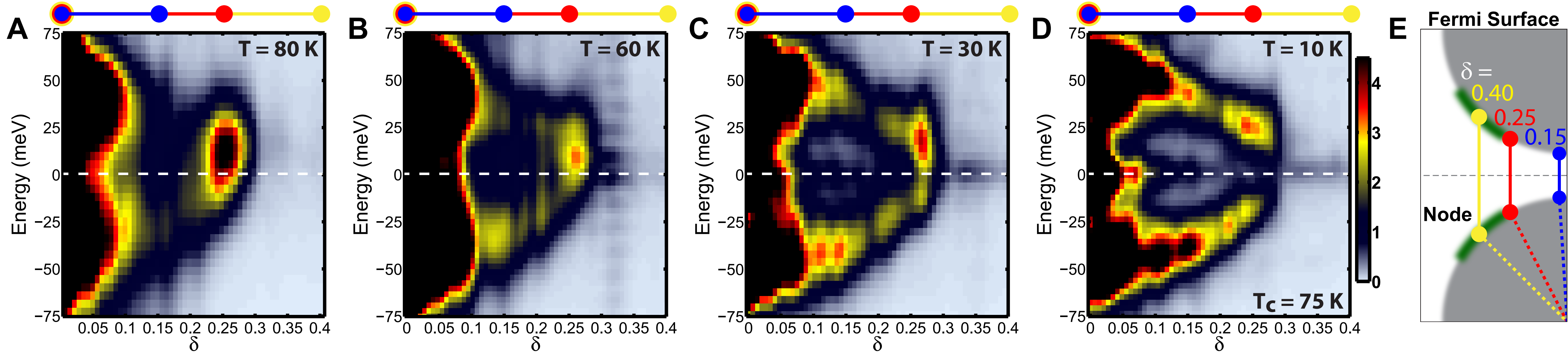

# Supplementary Materials for "Ubiquitous Interplay between Charge Ordering and High-Temperature Superconductivity in Cuprates"

Eduardo H. da Silva Neto[1*†], Pegor Aynajian[1*‡], Alex Frano[2,3], Riccardo Comin[4], Enrico Schierle[3], Eugen Weschke[3], András Gyenis[1], Jinsheng Wen[5], John Schneeloch[5], Zhijun Xu[5], Shimpei Ono[6], Genda Gu[5], Mathieu Le Tacon[2], and Ali Yazdani[1§]

[1]Joseph Henry Laboratories and Department of Physics, Princeton University, Princeton, New Jersey 08544 USA
[2]Max-Planck-Institut für Festkörperforschung, Heisenbergstrasse 1, D-70569 Stuttgart, Germany
[3]Helmholtz-Zentrum Berlin für Materialien und Energie, Albert-Einstein-Strasse 15, D-12489 Berlin, Germany
[4]Department of Physics & Astronomy, University of British Columbia, Vancouver, British Columbia V6T 1Z1, Canada
[5]Condensed Matter Physics and Materials Science, Brookhaven National Laboratory, Upton, NY 11973 USA
[6]Central Research Institute of Electric Power Industry, Komae, Tokyo, Japan

[*]These authors contributed equally to this work.
[†]Present address: Quantum Matter Institute, University of British Columbia, Vancouver, British Columbia V6T 1Z4, Canada
[‡]Present address: Department of Physics, Applied Physics and Astronomy, Binghamton University, Binghamton, NY 13902, United States
[§]To whom correspondence should be addressed. Email: yazdani@princeton.edu

**Data Acquisition and Analysis**

The STM measurements were carried out on a homebuilt variable temperature ultra-high vacuum scanning tunneling microscope. The conductance maps are measured using standard lock-in techniques and always using the same set-point bias of -300 mV. This value of set-point bias was carefully chosen to avoid systematics originating from the so-called *set-point effect* (see below). We typically chose a lock-in modulation of 3 mV for all measurements taken below 60 K and a set-point current between 150 and 200 pA. Measurements at 80 K were performed with a lock-in modulation of 4 mV (still much lower than thermal broadening) and currents between 60 and 100 pA. This was necessary due to the intrinsic instabilities of the tunneling junction on Bi-2212 at these high temperatures.

The conductance maps are taken on a square area equivalent of 20 b-axis supermodulations on a 256 x 256 pixel grid. We reject signals in the conductance maps that are two standard deviations away from the average of the map. At these pixels, we use the average value of the map. However, at the typical quality of the maps acquired for this work, less than 0.1% of pixels are rejected. Prior to carrying out the discrete Fourier transform (DFT), each conductance map is corrected for picometer-scale drifts by using the algorithm outlined by Lawler et al. (*30*). We use this drift correction to achieve the highest momentum resolution. However, even in the absence of the drift correction procedure, the typical deviations are smaller than the momentum-width associated to the charge ordering (CO) peak or other errors due to tip geometries.

Prior to performing the DFT of a conductance map, we normalize the maps by their mean to eliminate any dependence on junction impedance and to allow a direct comparison of the modulation's strength for different energies (also see Fig. S1). After carrying out the DFT, we use the mirror symmetry (relative to the a-axis of the crystal) to further suppress the influence of random tip geometries (which might have a preferential direction) on our measurements. Still, variations in the sharpness of the tip between different measurements can lead to large differences in the intensity of the CO feature in the DFTs. However, since we are interested only in the relative strength between the CO and BdG-QPI features, each energy-wavevector (E-$\delta$) plot is normalized to its average intensity in the area encompassed by -75 meV < E < 75 meV and ($\delta$* - 0.15) < $\delta$ < ($\delta$* + 0.05), where $\delta$* is the location of the CO feature (e.g. $\delta$*=0.3 for UD35). This choice of normalization area contains both CO and QPI features and allows a direct comparison of the interplay between CO and superconductivity for measurements done on different samples and at different temperatures. All the E-$\delta$ plots of the modulations along the Cu-O bond direction displayed in the text are produced by the procedure outline above.

**Influence of set point bias on charge ordering modulations**

In STM, a differential conductance map at a pre-set current $I_S$ and pre-set bias $V_S$ is given by

$$\frac{dI}{dV}(r, z, \omega = eV) = \frac{e\, I_S\, N(r,\omega)}{\int_0^{eV_S} N(r,\omega)\, d\omega} \tag{S1}$$

Where $e$ is the charge of the electron, $N(r, \omega)$ represents the electronic density of states at the spatial position $r$ and energy $\omega$. The denominator in Eq. S1 corresponds to the local density of states integrated within the Fermi energy and the set-point bias and represents what is known as the STM set-point effect. Since $N(r, \omega)$ is expected to vary locally, the denominator of Eq. S1 is also expected to vary locally. In most cases the spatial dependence of $N(r, \omega)$ follows the topograph and the effects of the set-point mirror in the atomic lattice (Bragg peaks in the DFTs). However, if strong modulations with an arbitrary wavevector exist at a specific energy $\omega = \omega_1$ lying within the set-point bias, then these modulations could show up in the experimentally measured $\frac{dI}{dV}(r, z, eV \neq \omega_1)$ for a wide range of energies through the denominator of Eq. S1. In other words, a modulation of the real density of states of the material at $\omega_1$ could appear at energies $\omega_2 \neq \omega_1$ in the *dI/dV* measurement (*31*). Since the denominator of Eq. S1 is controlled by the set-point bias e$V_s$, we can probe the effect of the set point on the measured charge ordering signal.

Figure S1 A-C shows the E-$\delta$ structure of the charge ordering obtained directly from the DFTs for a UD35 sample measured at 30 K, for different set-point biases. The central difference among the different datasets is the non-dispersive feature near $\delta = 0.3$. At $V_s = +200$ mV there are three branches of non-dispersive modulations near $\delta = 0.3$: E

> +75 mV, E < -50 mV and 0 < E < 50 mV. At $V_s$ = +400 mV the high energy branches become weaker, suggesting that the high energy branches are systematics of the measurement. Notice, however, that the low-energy branch (and 0 < E < 50 mV) seems unaltered by the change in $V_s$, indicating that it represents a real modulation of $N(\mathbf{r},\omega)$. Since the low-energy branch is only present above the chemical potential, it should not influence $\int_0^{eV_s} N(\mathbf{r},\omega)\,d\omega$ for $V_s < 0$. As a control experiment, in Fig. S1C we plot the same data measured with $V_s$ = -300 mV. The observation of only the low-energy 0 < E < 50 mV, with no shadows, confirms that the non-dispersive δ = 0.3 modulations exist only above the Fermi energy within 0 < E < 50 mV. Since it is expected that features associated with set-point effect systematics to be weaker than real modulations of $N(\mathbf{r},\omega)$, we further normalize each energy-specific real space conductance map to its spatially averaged value prior to the DFT operation. Figures S1 D-E, produced through this analysis, clearly show that the low energy branch is the dominant feature and therefore reinforces our identification of this feature as real and the other branches as systematics.

It is important to also note that the denominator in Eq. S1 can be removed by another post-processing analysis which has been widely used in the STM literature, the so-called Z-map, or Z-ratio. It is constructed as a ratio of positive and negative conductance maps, therefore canceling the denominator which contains the shadow signals.

$$Z(\mathbf{r},V) = \frac{dI/dV(\mathbf{r},\omega=+eV)}{dI/dV(\mathbf{r},\omega=-eV)} \tag{S2}$$

Indeed, Figs. S1 G-I show that all the high energy branches disappear in the Z-map. Unfortunately however, it can also be seen that the Z-map analysis is highly detrimental to the finer features observed in the raw data. For example, quite obviously, the information regarding particle-hole symmetry is lost in this analysis.

Altogether, Fig. S1 shows that the δ = 0.3 feature is a *low-energy* particle-hole asymmetric feature, absent at negative energies, and therefore must be measured with a choice of negative set-point bias. All the measurements shown in the main paper and below were performed with $V_s$ = -300 mV.

**Samples**

To achieve low doping levels (UD15, UD35, and UD45) the Bi-2212 samples were doped with Dy. To verify that the doping evolution seen in Fig. 2 of the main text is not affected by the presence of absence of Dy, we have taken the UD45 sample and oxygenated it at high pressure, resulting in the UD75 sample. Comparing the E-$\delta$ structure at 30 K between UD73 (no Dy, Fig. 2B) and UD75 (with Dy, Fig. 4C) shows that the presence of Dy has no influence on our STM measurements. Furthermore, notice that the same sample can show a CO feature with $\delta = 0.30$ (UD45) or $\delta = 0.25$ (UD75) depending solely on its oxygen content.

**X-ray: Methods and Analysis**

The REXS experiments reported here were performed with photon energies near the Cu-$L_3$ absorption edge (931.5 eV) which is resonant to the $2p_{3/2}$ -> 3d transitions, at the UE46-PGM1 beam line of the Helmholtz-Zentrum Berlin at BESSY-II. To maximize intensity, vertically polarized photons in a horizontal scattering geometry were used (see Fig. 1A of the main text). The 2-circle diffractometer was equipped with a continuous flow He cryostat reaching a base temperature of 10 K. The $\delta$ values reported here represent the component of momentum-transfer parallel to the Cu-O planes, obtained from theta scans at fixed detector angle 167°.

Figure S2 shows the temperature-dependent enhancement near $\delta = 0.29$. It can already be seen in the raw data that this peak evolves most rapidly for temperatures between 100 K and 200 K. The shape of the background (300 K) is indeed similar to all the previous YBCO measurements performed in this scattering chamber (*23*), although perhaps less apparent in that case due to the higher signal-background ratio of the YBCO CO peak. Particularly on resonance, the background's shape is a result of an intricate combination of systematic effects, including most prominently fluorescence, making it difficult to interpret quantitatively. On a qualitative account, the rocking curve scans we performed span 50°, so variations of 1 to 2% in the background are expected. Particularly on the far side (theta ~ 130°, two-theta=167°), grazing angles for both the incoming and take-off beam approach the sample horizon, causing the strong decrease of the background signal at large δ. Therefore due to the small amplitude of the signal-to-background ration, it is critical to properly characterize the background in the rocking scans.

Because the background can be dependent on the position of the beam on the sample, the sample position was adjusted at each temperature in order to correct for thermal drift and therefore maintain the beam position on the sample constant throughout all temperatures. Nevertheless, due to small changes in the background as a function of temperature, the fitting procedure outlined in Fig. S3 is used to extract the CO peak from the background. A fourth-order polynomial fit is performed on the tails (δ < 0.22 and δ > 0.37) and subtracted from the total signal in order to isolate the CO peak. After subtraction from the fitted background, we obtain Fig. S3B. To extract the temperature dependence of the peak, each data is fit to a Lorentzian function, resulting in Fig. 1F of the main text.

Figure S4 shows measurements performed at 10 and 300 K for negative values of δ (while maintaining the detector at the same fixed angle of 167° as was the case for positive δ). Contrasting the low and high temperature data clearly shows the presence of the CO peak. After background subtraction (subtraction of the high temperature data from the low temperature data) we obtained the peak in the inset of the Fig. S4. The δ value of the peak extracted from the fits is measured to be 0.28(1) from the data in Fig. S3B and 0.30(1) from Fig. S4 (δ<0). The determination from the STM is 0.30(1), in agreement with the diffraction data.

Figure S5 displays the dependence of the CO feature on the x-ray energy showing the enhancement near the resonance condition at the Cu-$L_3$ edge (931.5 eV).

**Analysis of particle-hole asymmetry in the STM data**

To better visualize the particle-hole asymmetry of the charge modulation in Bi-2212, in Fig. S6 we show the intensity of the E-δ structure in the DFT, integrated over positive ([0 to 50] meV blue curves) and negative ([0 to -50] meV red curves) energies, as a function of momentum for all the different samples. Depending on the momentum, the data contains information from both BdG-QPI and CO modulations. Remarkably, all samples show a striking particle-hole asymmetry only in the momentum range around δ=0.3 for UD15, UD35, UD45 and δ=0.25 for UD52, UD61, UD73, corresponding to the charge ordering phenomena. The optimal doped sample, OP91, displays particle-hole symmetry over the entire momentum range, expected from the BdG-QPI. A blow-up of the data around the critical momenta for all the samples is shown in Fig. S6 B.

Figure S6 C displays energy cuts of the intensity of the E-δ structure for two constant momenta δ = 0.2, corresponding to a BdG-QPI wavevector and δ = 0.30 (0.25) corresponding to the critical charge ordering wavevector. For δ=0.2, the data show a particle-hole symmetry for positive and negative energies as expected from BdG-QPI, whereas for the critical CO wavevector, the data show a clear asymmetry indicating that the CO exist above the Fermi energy in all these samples.

Figure S7 displays similar data to Fig. S6 for a UD75 sample as a function of temperature across $T_c$. The data clearly shows the evolution of the particle-hole asymmetry as a function of increasing temperature.

**Doping dependence of CO energy and δ-value**

The central energy of the CO resonance remains above chemical potential (approximately +20 meV) for all samples with $T_c$ values between 15 K and 73 K. This constancy can already be seen in Fig. 2B-G, and is summarized in Fig. 2H. The energy values displayed in Fig. 2H correspond to the center of the CO feature and were obtained by fitting the positive part of the curves in Fig. S6C to a Lorentzian function. The error bars displayed in Fig. 2H for the energy values correspond to the uncertainty obtained from the least squares fitting (95% confidence interval).

Figure 2H of the main text displays a sudden jump in the δ-value of the CO: δ ≈ 0.30 for UD45 and δ ≈ 0.25 for UD52. This jump can already be seen directly in analysis presented in Fig. S6 A-B. For each sample the energy integrated δ-dependent curves are fit to a Lorentzian function plus a step like background. An example of this fitting procedure is illustrated in Fig. S8, for a UD45 sample. The δ value of the CO plotted in Fig. 2H corresponds to the location of the Lorentzian peak from the fit and the error bar is the half width at half maximum.

**Regions of the Fermi surface associated with charge order**

Figure 3 of the main text shows that the δ value of the charge order feature seen in the STM data is associated with the regions of the Fermi surface near the end of the Fermi arcs. In this section we describe in detail how the Fermi surface of Fig. 3E was constructed, and its relation to previously published ARPES data (*28,29*).

Figure S9 shows a schematic Fermi surface for a Bi-2212 UD75 sample. For simplicity, the hole barrels are modeled as circles, with their radii adjusted to match the nodes as determined by ARPES (*29*). The purple arc near the node is terminated at the angle where a deviation of the gap value from the $d_{x^2-y^2}$ functional form is observed in ARPES measurements below $T_c$ (*28*). The white segment near the node represents the Fermi arc above $T_c$, corresponding to zero gap regions, as seen by ARPES (*28*). Note that the white segment in Fig. S9 corresponds to the green segment in Fig. 3E.

Following this construction, the anti-nodal BdG-QPI scattering at high energies is expected to correspond to δ = 0.15 (see blue line in Fig. S9). This corresponds approximately to the lowest value of δ ≈ 0.12 detected in our STM measurements around ± 45 meV. Contrasting these measurements to high energies we find that the δ-value associated with CO connects regions of the Fermi surface lying on the Fermi arcs of the d-wave order parameter (see red line in Fig. S9).

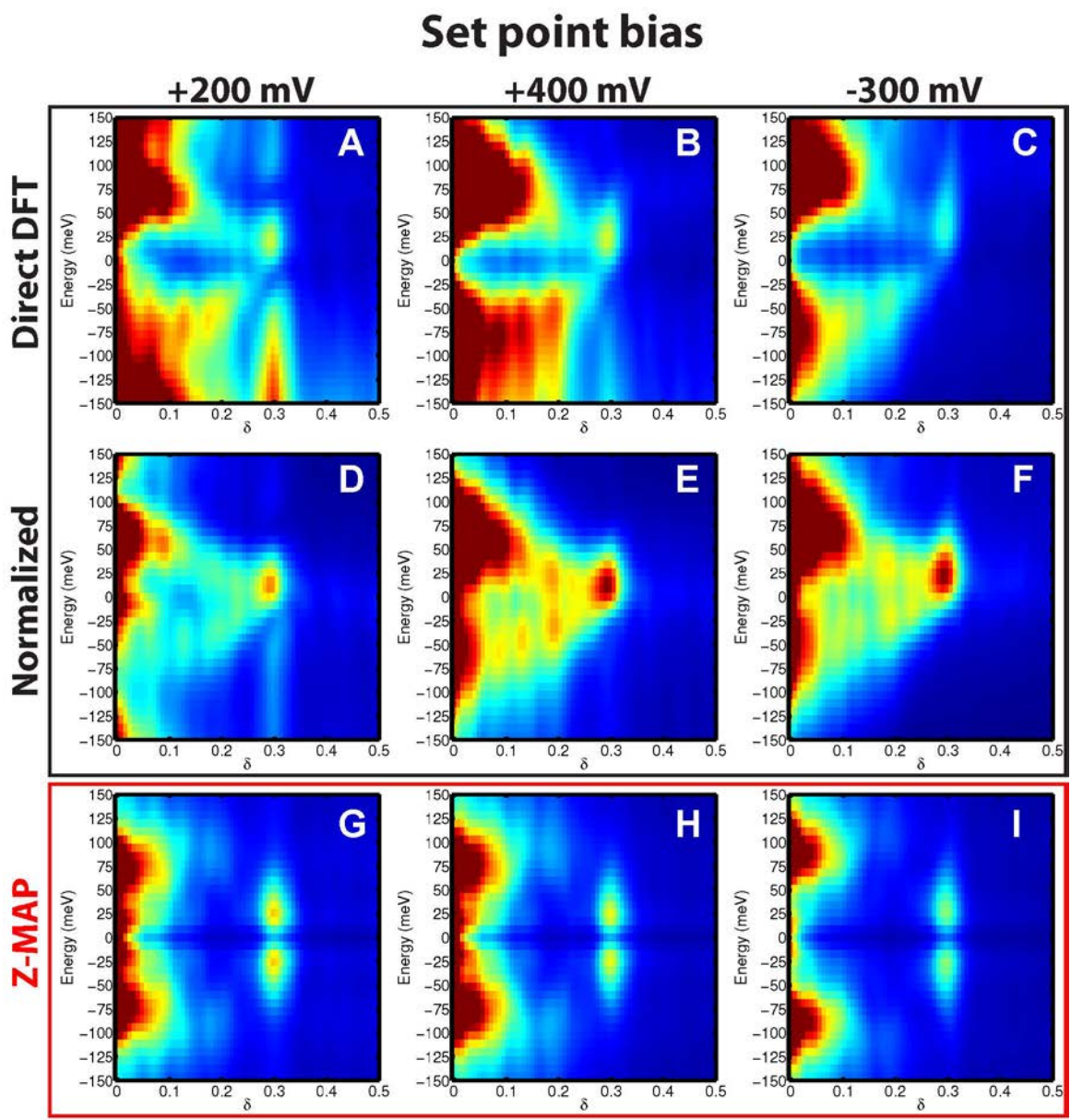

Fig. S1.

**A-C** E-δ structure of the LDOS modulations along the Cu-O bond direction obtained directly from the DFT of the conductance maps. **D-F** similar to **A-C** but with the real-space conductance maps normalized to the spatially averaged energy spectrum. **G-I** Z-maps constructed from Eq. S2. The Z-map construction was reflected about zero-energy to illustrate its loss of information regarding particle-hole symmetry.

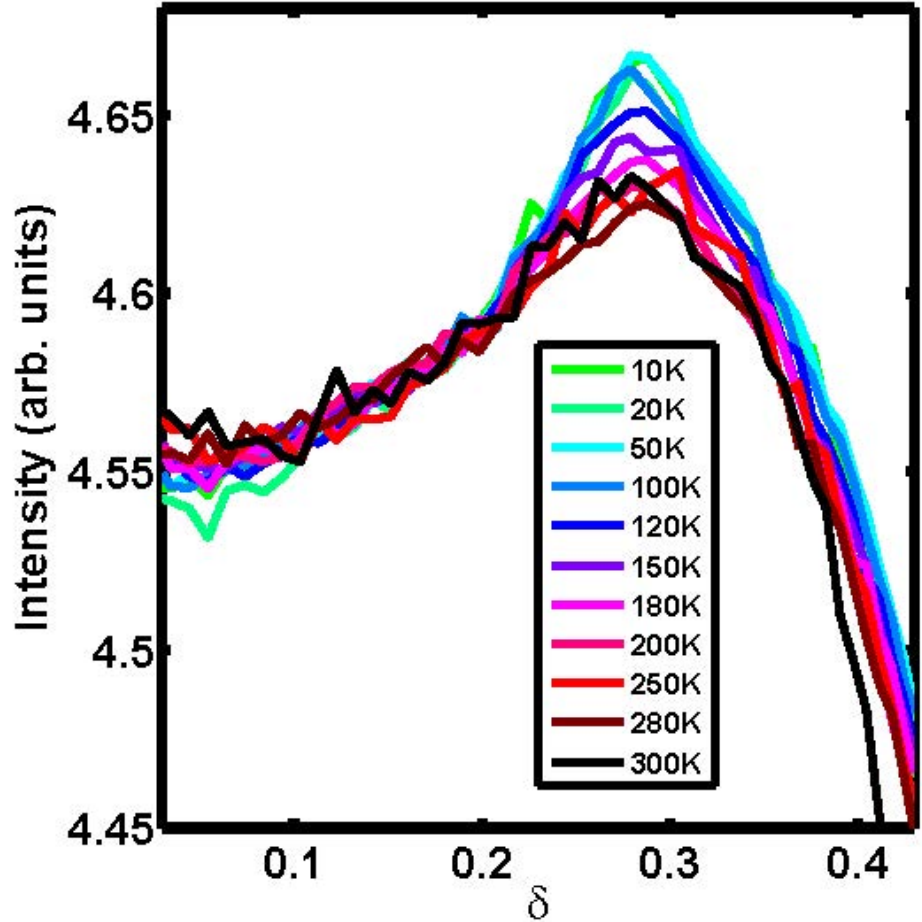

Fig. S2
For each temperature, theta scans were performed ten times and averaged to obtain a better signal-to-noise ratio. Each curve is shifted by a constant to account for a temperature dependent offset of the background.

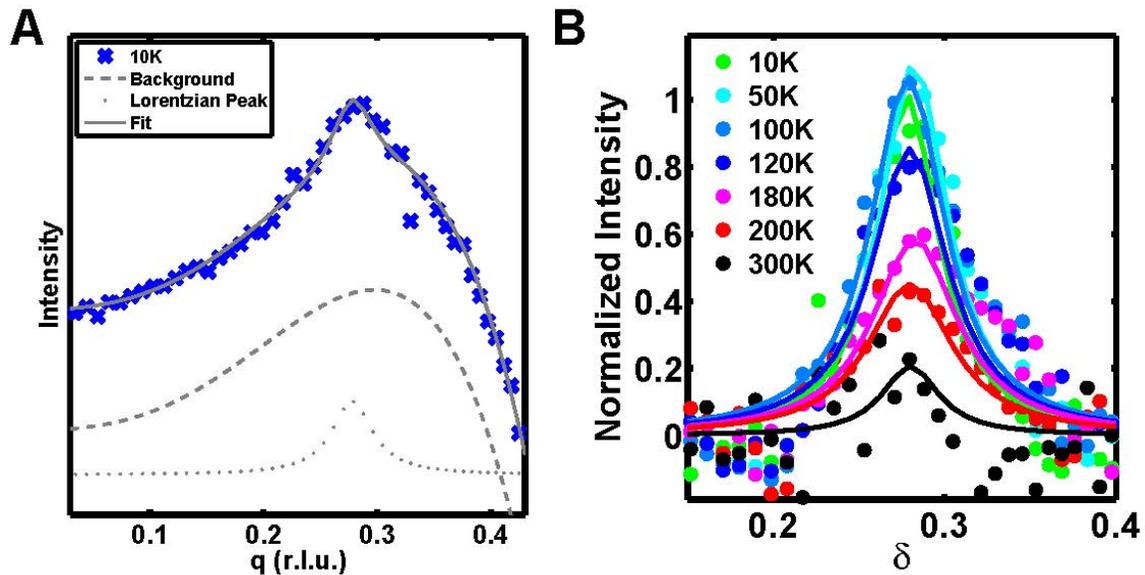

Fig. S3

**A** Background is fitted to a fourth-order polynomial for δ < 0.22 and δ > 0.37. A Lorentzian function is fitted to the data after background subtraction. Notice the small kink in the raw data above δ = 0.22 signaling the presence of the peak on top of the smooth background. **B** Background subtracted REXS δ-scans (for positive δ) for a UD45 sample at selected temperature, showing the evolution of the CO peak near δ = 0.3. The lines are Lorentzian fits to the data.

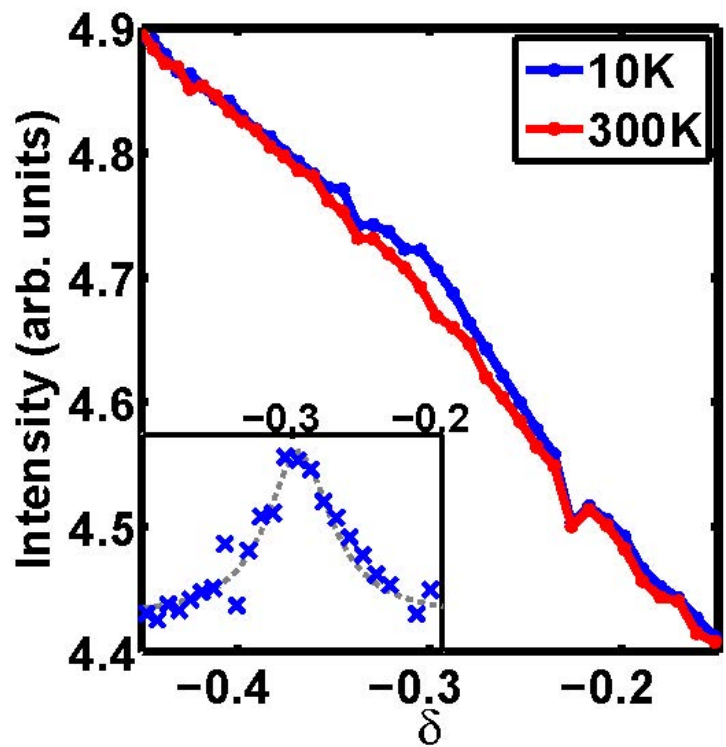

Fig. S4
REXS δ-scans for negative δ at low (blue) and high (red) temperatures, showing a clear enhancement near δ = 0.3 at low temperatures. Inset shows the background subtracted 10 K data, where the dashed line represent a Lorentzian fit to the data.

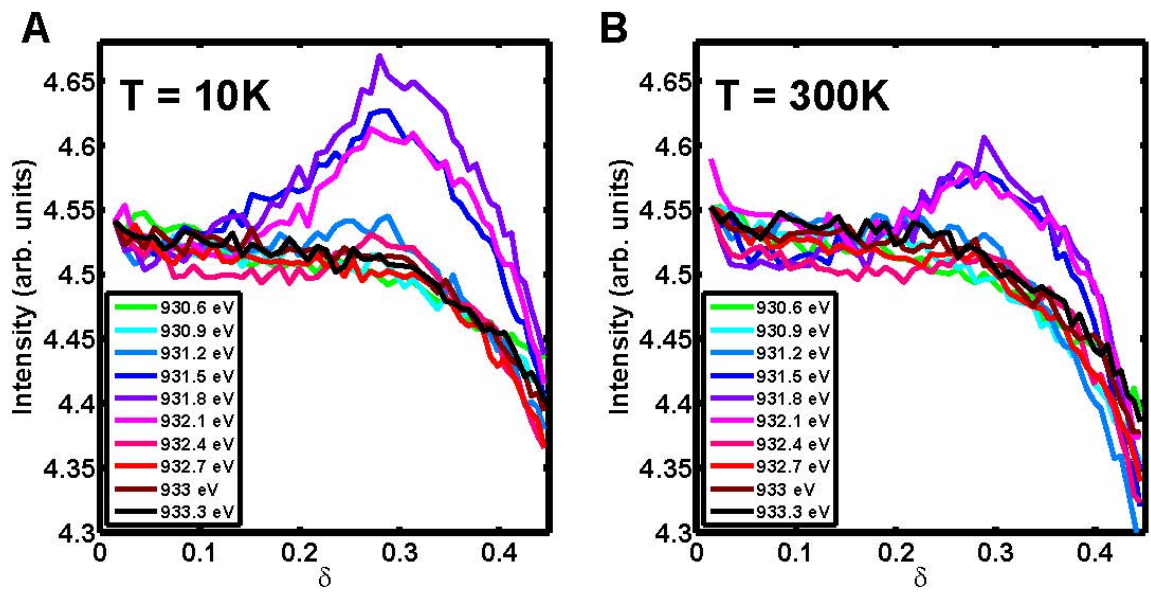

Fig. S5
The dependence of δ > 0 peak on photon energy near the Cu-$L_3$ edge (931.5 eV), at 10K and 300K. For each energy, theta scans were performed three times and averaged to obtain a better signal-to-noise ratio. Each curve is shifted by a constant to account for an energy dependent offset of the background.

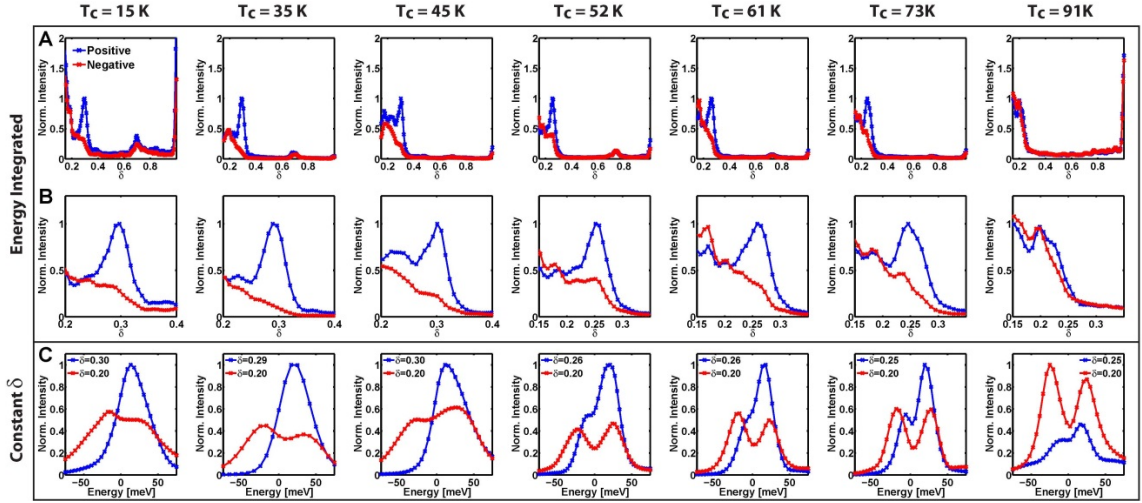

Fig. S6

**A** E-δ structure in the DFT, integrated over positive ([0 to 50] meV blue curves) and negative ([0 to -50] meV red curves) energies, as a function of momentum for all the different samples. **B** Same as in **A** but in a narrower range around δ = 0.3 for UD15, UD35, UD45 and δ = 0.25 for UD52, UD61, UD73, OP91. All the curves in A and B are normalized to their maximum near δ = 0.3 or δ = 0.25 accordingly. **C** Energy cuts of the intensity of the E-δ structure for two constant momenta δ = 0.2 and δ = 0.30 (0.25). Curves were normalized to maximum of the two curves.

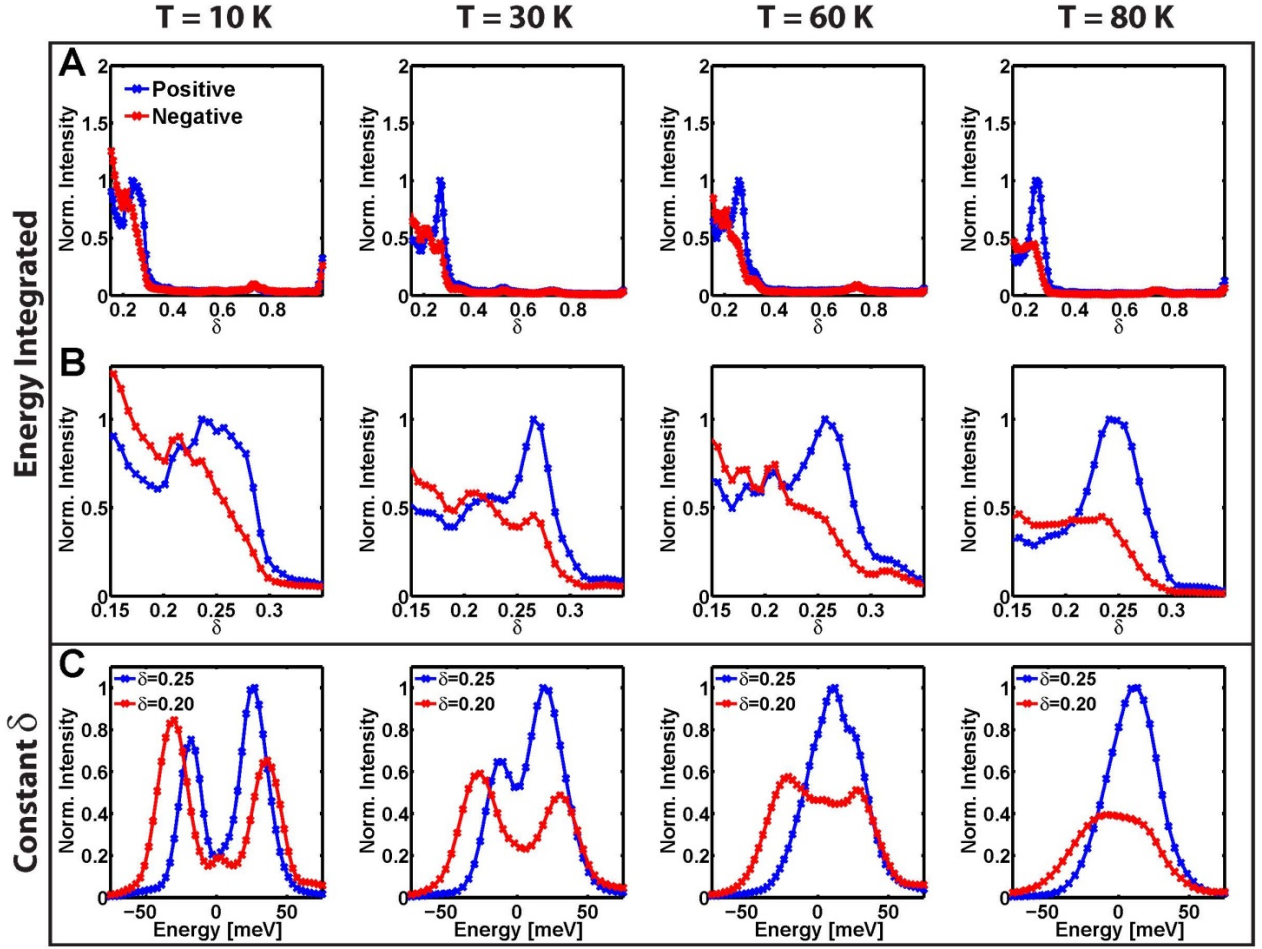

Fig. S7
**A** E-δ structure in the DFT, integrated over positive ([0 to 50] meV blue curves) and negative ([0 to -50] meV red curves) energies, as a function of momentum for all the different samples. **B** Same as in **A** but in a narrower range around δ = 0.25. All the curves in **A** and **B** are normalized to their maximum near δ = 0.3 or δ = 0.25 accordingly. **C** Energy cuts of the intensity of the E-δ structure for two constant momenta δ = 0.2 and δ = 0.25. Curves were normalized to maximum of the two curves.

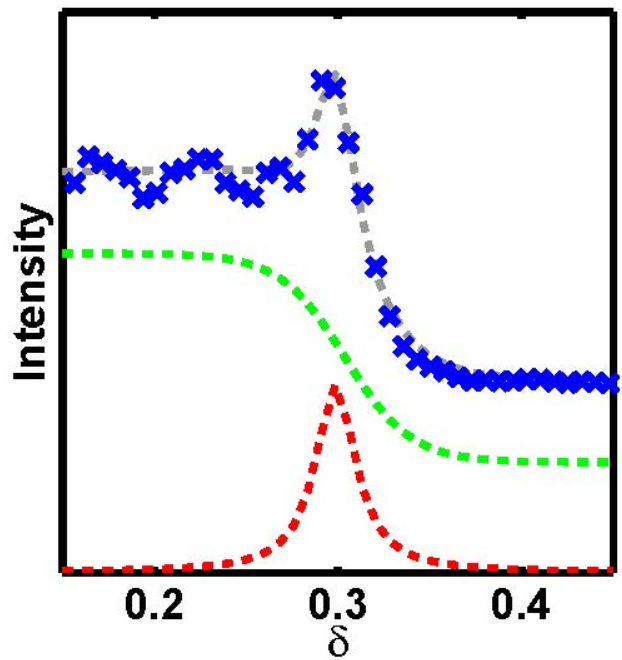

Fig. S8
The energy integrated (0 to 50 meV) δ-cut (blue crosses) is fitted to a smooth step-like background (green) plus a Lorentzian function (red). Curves are offset for clarity.

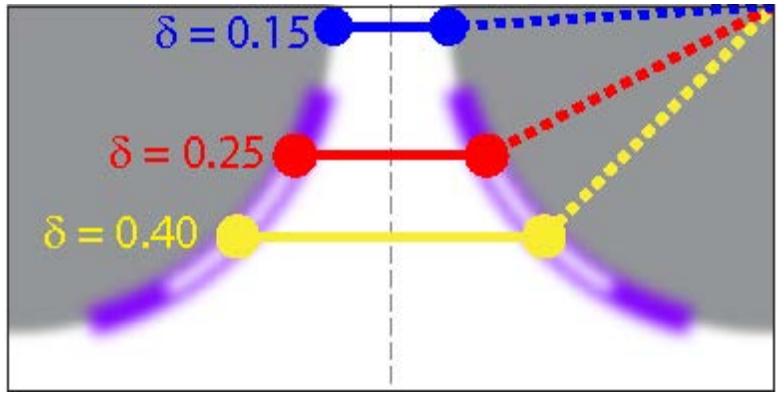

Fig. S9
Schematic of the Fermi surface of Bi-2212 for a UD75 sample showing the value of $\delta = 0.15$ for anti-nodal scattering (blue line). The red line represents the value of $\delta = 0.25$ associated to the CO in this sample. The hole barrels (dark gray) are modeled as circles, with their radii adjusted to match the location of the nodes as determined by ARPES measurements on similar samples (*29*).